# Ordinal Classification Method for the Evaluation Of Thai Non-life Insurance Companies

**Phaiboon Jhonpita[1], Sukree Sinthupinyo[2] and Thitivadee Chaiyawat[3]**

[1]Technopreneurship and Innovation Management Program Graduate School, Chulalongkorn University,
Bangkok, Thailand
phaiboonj@gmail.com

[2]Department of Computer Engineering, Faculty of Engineering, Chulalongkorn University,
Bangkok, Thailand
sukree.s@ chula.ac.th

[3]Department of Statistics, Faculty of Commerce and Accountancy Chulalongkorn University,
Bangkok, Thailand
Thitivadee@acc.chula.ac.th

**Abstract**

This paper proposes a use of an ordinal classifier to evaluate the financial solidity of non-life insurance companies as strong, moderate, weak, and insolvency. This study constructed an efficient classification model that can be used by regulators to evaluate the financial solidity and to determine the priority of further examination as an early warning system. The proposed model is beneficial to policy-makers to create guidelines for the solvency regulations and roles of the government in protecting the public against insolvency.

## 1. Introduction

Thailand Insurance industry is subject to government regulation to protect policyholders, third-party liability claimant, and other related business. Solvency supervision, regulations and solvency position classification is an important topic for non-life insurers. Most of the studies were implemented in the United States and many previous studies focused on binary classification and the problem whose class values were unordered (bankrupt/non-bankrupt, solvency/insolvency, or healthy/failed)[2-16]. Unfortunately, they were not implemented in the multi-class classification fashion. In this paper, we hence proposed an ordinal multi-class classification for solvency condition classification. Normally, The Office of Insurance Commission (OIC) of Thailand uses the Capital ratio (CAR) system of non-life insurance in 2009 to evaluate the capital adequacy or financial solidity of the non-life insurers (as shown in Table 1). with the condition distinguished by a level of CAR, the insurance company and regulator's actions are required.

**TABLE 1** The solvency evaluation and regulatory actions based on CAR system.

| Class Classification | Capital adequacy ratio (CAR) | The action level |
|---|---|---|
| Strong | ≥ 150% | No action level |
| Moderate | 120 - 150% | Company action level |
| Weak | 100 - 120% | Regulatory action level |
| Insolvency | < 100% | Authorized control & Mandatory control level |

Note: Company action level - company must file plan with insurance commissioner & explaining cause of deficiency and how it will be corrected. Regulatory action level - The commissioner is required to examine the insurer and take corrective action, if necessary. Authorized control level & Mandatory control level - The commissioner has legal grounds to rehabilitate or liquidate the company, the commissioner is required to seize a company.

The level of capital adequacy ratio (CAR) of insurer is affected by most insurance activities and decision making processes such as premium rate making, determination of the technical reserve, risk undertaking, reinsurance activities, investment, sales, credibility of company to related party, and also be affected by the country's economy, new legislations, inflation and interest rates [1]. With the help of our system, the companies can early detect the solvency condition of their own and can decide the most suitable policy to reduce their risk.

## 2. Literature review

Among many empirical studies of insurance science, there are several studies with different techniques used for improving the performance of Insolvency prediction and/or classification model. Most studies applied traditional statistic techniques, such as regression analysis [2], multivariate discriminant analysis (MDA) [3, 4, 5], logistic regression (LR) [6], logit and probit model [7-10], and multinomial logistic regression (MLR) [1]. On the other hand, machine learning techniques such as neural networks (NNs) [11-15], and genetic algorithm (GA) [16] were also used in Insolvency prediction.

Kramer (1997) evaluated the financial solidity of Dutch non-life insurance by combining a traditional statistic technique (ordered logit model) with artificial intelligence techniques (a neural network and an expert system). The complete model contains three programs; logit model, neural network, and expert system. The data from year 1992 has been used as training data set and year 1993 as the test set. The output of the multi-class classification model consists of the priority for further examination (High, Medium, and Low class). The system which evaluates the financial solidity can be used to classify the insurers according to their degree of risk exposures. The model correctly classified 93% of the data test set. It showed very good performance for strong, medium and weak companies, 96.3% of the strong, 75.0% of the medium and 94.4% of the weak are classified correctly.

Pitselis (2009) studied the solvency supervision, regulations and insolvency prediction of Greece insurance companies using statistical methodologies, e.g. discriminant analysis (DA), logistic regression (LR), and multinomial logistic regression (MLR) to distinguish solvency position into two cases; two-class classification (healthy and insolvency) and multi-class classification (healthy, merged, and insolvency). The paper presented the effects of solvency position of insurance companies. Company and regulatory actions are required if a company's solvency position falls below requirement. Due to the imbalanced data problem, especially for insolvency companies, LR and MLR failed to give reliable results. DA model was able to adequately classify Healthy, Merged, Insolvency companies; 93.5%, 33.3% and 100% respectively (on the 1998 data set).

### 2.1 A Simple Approach to Ordinal Classification

Frank and Hall (2001) [17] presented an ordinal classification approach that enables standard classification algorithms to classify the ordinal class problems. Frank and Hall applied standard classifier in conjunction with a decision tree learner. The underlying learning algorithm takes advantage of ordered class values. First, the original dataset problem is transformed from a $k$-class $V = \{v_1 \ldots v_k\}$ to $k-1$ binary-class problems. The training starts by deriving new datasets from the original dataset, one for each of the $k$-1 new class attributes. In the next step, the classification algorithm is applied to generate a model for each of the new datasets. To predict the class value of an unseen instance, we need to estimate the probabilities of the $k$ original ordinal classes using our $k$-1 model. Estimation of the probability for the first and last ordinal class value depends on a single classifier.

In General, for class values $V_i$, a probabilities distribution on $V_i$ ($k$-classes) is then derived as follows:

$Pr(V_1) = 1 - Pr(\text{Target} > V1)$
$Pr(V_i) = max\{Pr(\text{Target} > V_{i-1}) - Pr(\text{Target} > V_i), 0\}, 1 < i < k$
$Pr(V_k) = 1 - Pr(\text{Target} > V_{k-1})$

To classify an instance of an unknown class, each of the k-1 classifiers and the probabilities of each the k ordinal class value is calculated using method above evaluate the instance. The class with maximum probability is assigned to that instance.

### 2.2 Decision Tree Learning Algorithm

The Decision Tree Learning (DTL) algorithm we used in this research is the one named J48 implemented in WEKA machine learning tool [18]. The J48 class is implemented based on the same concept as C4.5 decision tree [19].

The DTL is a predictive machine learning model which begins with a set of the whole training examples. It creates a decision tree based on the attribute values of the training data that can best classify the set of samples at a time. The attribute which can best discriminate the sample set is evaluated based on the concept of Entropy. The examples are then divided into edges which is the value of the attribute. The child node which consists of examples from different classes will be replaced with the new attribute node, while the child node containing examples from the same class will be a used as a decision node, in which all examples will be classified as the class of training examples collected in this node.

## 3. Data and Methodology

The data set used in this study was collected from 70 non-life insurance companies in Thailand. The companies which were in operation or went insolvency were covered from 2000 to 2008. During this period, 616 cases (543 strong, 16 moderate, 13 weak and 44 insolvency) were selected as training data set as shown in Table 2. The data of year 2009 were used as a separated test set. The data source comes from the annual report of The Office of Insurance Commission (OIC) and the health insurance companies are not including on this study.

**TABLE 2** Number of Non-life Insurance companies in this study (Data from year 2009 are the separated test set).

| Class | 2000 | 2001 | 2002 | 2003 | 2004 | 2005 | 2006 | 2007 | 2008 | Total | % | 2009 |
|---|---|---|---|---|---|---|---|---|---|---|---|---|
| Insolvency | 5 | 3 | 6 | 5 | 7 | 4 | 6 | 5 | 4 | 45 | 7.1% | 6 |
| Weak | 1 | 1 | 1 | 1 | 2 | 0 | 3 | 1 | 3 | 13 | 2.1% | 1 |
| Moderate | 0 | 1 | 1 | 2 | 2 | 4 | 3 | 3 | 1 | 17 | 2.6% | 1 |
| Strong | 64 | 65 | 62 | 62 | 59 | 60 | 56 | 56 | 57 | 541 | 88.1% | 57 |
| Total | 70 | 70 | 70 | 70 | 70 | 68 | 68 | 65 | 65 | 616 | 100 % | 65 |

Note: The solvency condition in this study is determined by capital adequacy ratio
= Total capital available (TCA) / Total capital required (TCR)

The attributes selection started from 13 attributes. We chose them from the most commonly used ones in empirical studies of insurance science. They were found significant in previous studies of predicting non-life insurances' solvency [1-11, 13-16]. In this paper, we select the relevant attributes using the correlation-based attribute subset evaluator and greedy stepwise. All 13 attributed are shown in Table 3.

**TABLE 3** Attributes used in this study

V1  Net premiums written / policyholders' surplus
V2  Solvency margin to minimum required solvency margin
V3  Policyholders' surplus & Technical reserve to net written premium
V4  Claims incurred to policyholders' surplus & technical reserve
V5  Gross agent's balance to policyholders' surplus
V6  Change in policyholders' surplus
V7  Investment yield
V8  Investment assets to policyholders' surplus
V9  Return on total assets (ROA)
V10 Loan & other investment to policyholders' surplus
V11 Loss reserve & unpaid losses to policyholders' surplus
V12 Capitalization ratio
V13 Auto lines net written premium to total net written premium

After we analyzed the distribution of the training data, we found that the distribution of the data set was imbalanced, as shown in Table 2. The classification of data with imbalanced class distribution has posed a significant drawback on the performance of most standard classifiers, which assume a relatively balanced class distribution and equal misclassification costs [20]. Many techniques were proposed to solve this problem, for example, re-sampling methods for the balancing the data set, modification of existing learning algorithms, measuring the classifier performance in imbalance domains, relationship between class imbalance, and other data complexity characteristics [21].

To attack the imbalanced data set problem, we employ the standard resample technique to produce a new random set of data by sampling with replacement. The distribution on the data sets after applying resample techniques is presented in Table 4. In this study, we use the ordinal class classifier which employs the DTL algorithm as the base classifier.

Figure 1 shows the classification process. Fig. 2 and 3 shows the concept of testing approaches, 10 fold cross-validations and 70:30% split data set validation.

**TABLE 4** Training data set after applying resample technique.

| Class Classification | Original data set | | Resample data set | |
|---|---|---|---|---|
| Insolvency | 45 | 7.3% | 157 | 25.5% |
| Weak | 13 | 2.1% | 137 | 22.2% |
| Moderate | 17 | 2.8% | 144 | 23.4% |
| Strong | 541 | 87.8% | 178 | 28.9% |
| Total | 616 | 100.0% | 616 | 100.0% |

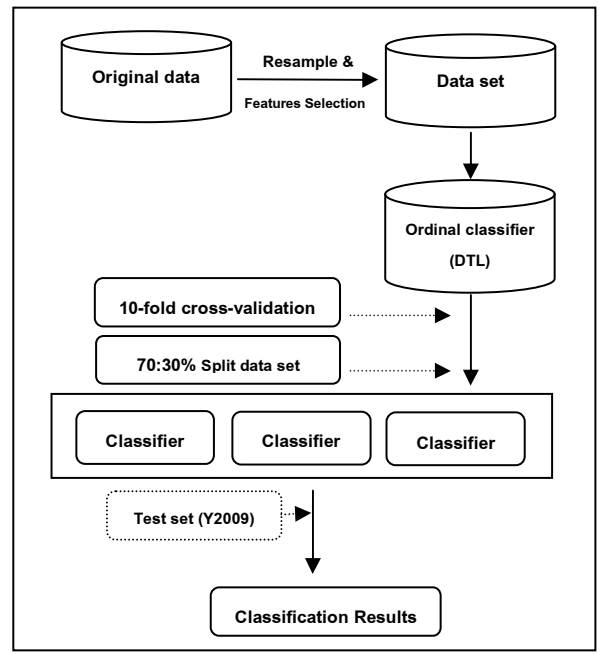

**Fig.1** Model Construction.

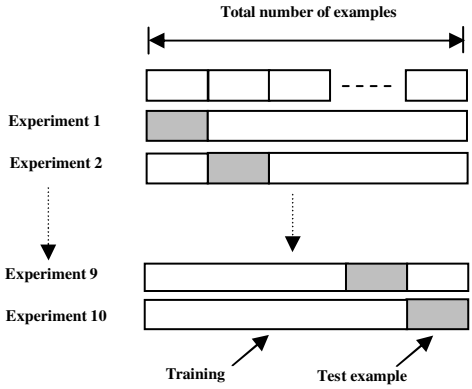

**Fig.2** 10-fold cross-validation

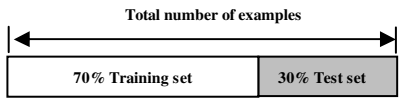

**Fig.3** 70:30% Split data set

## 4. Experimental and Results

This paper used a 10-fold cross-validation, 30% split test set and separated test set (2009 data set). The classification results are shown in Table 5, 6, and 7.

**TABLE 5** Classification results obtained from 10-fold cross-validation (total 616 instances)

| Class Classification | I | W | M | S | Total | Classified Correctly (%) |
|---|---|---|---|---|---|---|
| I | 154 | 3 | 0 | 0 | 157 | 98.1% |
| W | 0 | 137 | 0 | 0 | 137 | 100.0% |
| M | 0 | 0 | 144 | 0 | 144 | 100.0% |
| S | 0 | 0 | 5 | 173 | 178 | 97.2% |
| Total | | | | | 616 | 98.7% |

I = insolvency, W = weak, M = moderate, S = strong

**TABLE 6** Classification results from 30% spilt test set (total 185 instances)

| Class Classification | I | W | M | S | Total | Classified Correctly (%) |
|---|---|---|---|---|---|---|
| I | 49 | 2 | 0 | 0 | 51 | 96.1% |
| W | 0 | 44 | 0 | 0 | 44 | 100.0% |
| M | 0 | 0 | 40 | 3 | 43 | 93.0% |
| S | 0 | 1 | 2 | 44 | 47 | 93.6% |
| Total | | | | | 185 | 95.7% |

I = insolvency, W = weak, M = moderate, S = strong

**TABLE 7** Classification results from test set (2009 data set, 65 instances in total)

| Class Classification | I | W | M | S | Total | Classified Correctly (%) |
|---|---|---|---|---|---|---|
| I | 4 | 2 | 0 | 0 | 6 | 66.7% |
| W | 0 | 1 | 0 | 0 | 1 | 100.0% |
| M | 0 | 0 | 1 | 0 | 1 | 100.0% |
| S | 0 | 0 | 3 | 54 | 57 | 94.7% |
| Total | | | | | 65 | 92.3% |

I = insolvency, W = weak, M = moderate, S = strong

The results of applying the ordinal class classifier and DTL algorithms on the data introduced above depend on our selected financial ratios (attributes). The model shows a good performance and correctly classifies 98.7% from 10-fold cross-validation, 95.7% from 30% spilt test set, and 92.3% from the separated test set. The model can classify the minority class well but fail to recognize insolvency class in the separated test set (66.7% correctly classify). The relative importance of each attribute (input variable) is analyzed by calculating the weak class of the relationship between each input and output attribute.

**TABLE 8** Performance evaluation measure

| Cross-validation method | Evaluation | |
|---|---|---|
| | MAE | RMSE |
| 10 fold cross-validation | 0.0132 | 0.0838 |
| 30% spilt test set | 0.0281 | 0.1475 |
| Test set (2009 data set) | 0.0453 | 0.1985 |

MAE- Mean absolute error
RMSE- Root mean squared error

Table 8 presents performance evaluation measure of numeric prediction. In this study, we evaluated the performance of prediction by MAE and RMSE. The MAE and RMSE are given by

Mean absolute error (MAE)

$$= \frac{|p_1 - a_1| + \cdots + |p_n - a_n|}{n}$$

Root mean squared error (RMSE)

$$= \sqrt{\frac{(p_1 - a_1)^2 + \cdots + (p_n - a_n)^2}{n}}$$

Where, $P_1, P_2, ..., P_n$ denote the predicted values on the test instances and $a_1, a_2, ..., a_n$ denote the actual values.

## 5. Conclusions

From the experiment setting and results reported in the previous section, the results indicate that the obtained model can solve the problems of the multi-class classification and also the imbalanced data set. In this study, we employ the ordinal class classifier to solve the multi-class problem, so that our model can classify the solvency condition of Thai

Non-life insurance companies into four cases, strong, moderate, weak, and insolvency. To attack the problem of imbalanced data set, we use the standard resample technique which can highly improve the accuracy of the minority class which is the class that we are interested. Our final model are useful for insurance regulators, auditors, investors, management, policy holders, and related party to determine the priority for further examinations as an early warning system. In our further research, we will apply the ensemble methods and standard classifiers proposed here to better improve the imbalanced data set problem.